\documentclass[%
 reprint,prl
 amsmath,amssymb,
 aps,
]{revtex4-1}

\usepackage{graphicx}
\usepackage{multirow}
\usepackage{dcolumn}
\usepackage{bm}
\usepackage[left=1.2cm, top=2cm, right=1.2cm, textwidth=10cm]{geometry}
\usepackage{physics}
\usepackage[colorlinks=true
,urlcolor=blue
,anchorcolor=black
,citecolor=black
,filecolor=black
,linkcolor=black
,menucolor=black
,linktocpage=true
,pdfproducer=medialab
,bookmarks=false]{hyperref}
\usepackage{fullpage}
\usepackage{float}
\usepackage{latexsym}
\usepackage{epsf}
\usepackage{amsmath,amssymb,amsthm}
\usepackage[swedish, english]{babel}
\usepackage[utf8]{inputenc}
\usepackage{tikz}
\usepackage{enumerate}
\usepackage{verbatim}
\usepackage{stmaryrd}
\usepackage{mathtools}
\usepackage{tensor}
\usepackage[all]{xy}

\DeclareMathOperator{\R}{\mathbb{R}}
\DeclareMathOperator{\D}{\text{D}}
\DeclareMathOperator{\Ch}{\text{Ch}}
\DeclareMathOperator{\PT}{\mathcal{PT}}

\begin{document}

\title{Classification of Exceptional Nodal Topologies Protected by $\mathcal{PT}$ Symmetry}

\author{Marcus St\aa lhammar}
\author{Emil J. Bergholtz}
\affiliation{Department of Physics, Stockholm University, AlbaNova University Center, 106 91 Stockholm, Sweden}

\begin{abstract} Exceptional degeneracies, at which both eigenvalues and eigenvectors coalesce, and parity-time ($\PT$) symmetry, reflecting balanced gain and loss in photonic systems, are paramount concepts in non-Hermitian systems. We here complete the topological classification of exceptional nodal degeneracies protected by $\PT$ symmetry in up to three dimensions and provide simple example models whose exceptional nodal topologies include previously overlooked possibilities such as second-order knotted surfaces of arbitrary genus, third-order knots and fourth-order points. 

\end{abstract}

\maketitle

\section{Introduction} 
Parity-time ($\PT$) symmetric non-Hermitian (NH) descriptions were originally suggested as a fundamental amendment to the standard quantum mechanics motivated by their capacity of having real spectra \cite{benderboettcher,bender}. By now, NH models are instead recognized as effective descriptions with a remarkably wide range of applications \cite{NHbook}, and $\PT$ symmetry is an ubiquitous feature of photonic experiments where it reflects a balance between gain and loss \cite{OzRoNoYa2019}. Exceptional point (EP) degeneracies at which both eigenvalues and eigenvectors coalesce is a central and uniquely NH feature associated with a rich phenomenology in these systems \cite{EPreview}. Recently, the merger between topological \cite{hasankane,qizhang} and NH physics has led to the burgeoning field of NH topological systems \cite{review}.

The topological properties of NH nodal band structures, consisting of EPs, are central topic within this field \cite{review}. EPs are topologically distinguished from Hermitian degeneracies in that they are generic in the bulk of two-dimensional (2D) systems \cite{BerryDeg,Heiss,koziifu,NHarc}, giving closed (potentially knotted) lines of EPs in 3D \cite{EPrings,CaBe2018,EPringExp,ourknots,ourknots2,chinghuaknot,yangknot}. In general, EPs of order $n$ ($n$EPs) require the tuning of $2n-2$ real parameters, resulting in their absence for $n\geq 3$ in systems of 3D or lower. The inclusion of discrete symmetries can, however, relax this constraint and make symmetry-protected 2EPs generic already in one dimension (1D), while rings and surfaces of 2EPs are stabilized in 2D and 3D, respectively \cite{Budich2019,Yoshida2019,Okugawa,Zhou,HCphot,Etorus}. These EPs are associated with open Fermi arc degeneracies in the real part of the spectrum with telltale signatures in experiments \cite{NHarc}. Very recently, it was shown that physically relevant symmetries can also  stabilize higher-order EPs, and explicit examples of models with point-like 3EPs in 2D were explored \cite{3Tsuneya,ips3}.

\begin{figure}[hbt!]
\centering
\includegraphics[width=\columnwidth]{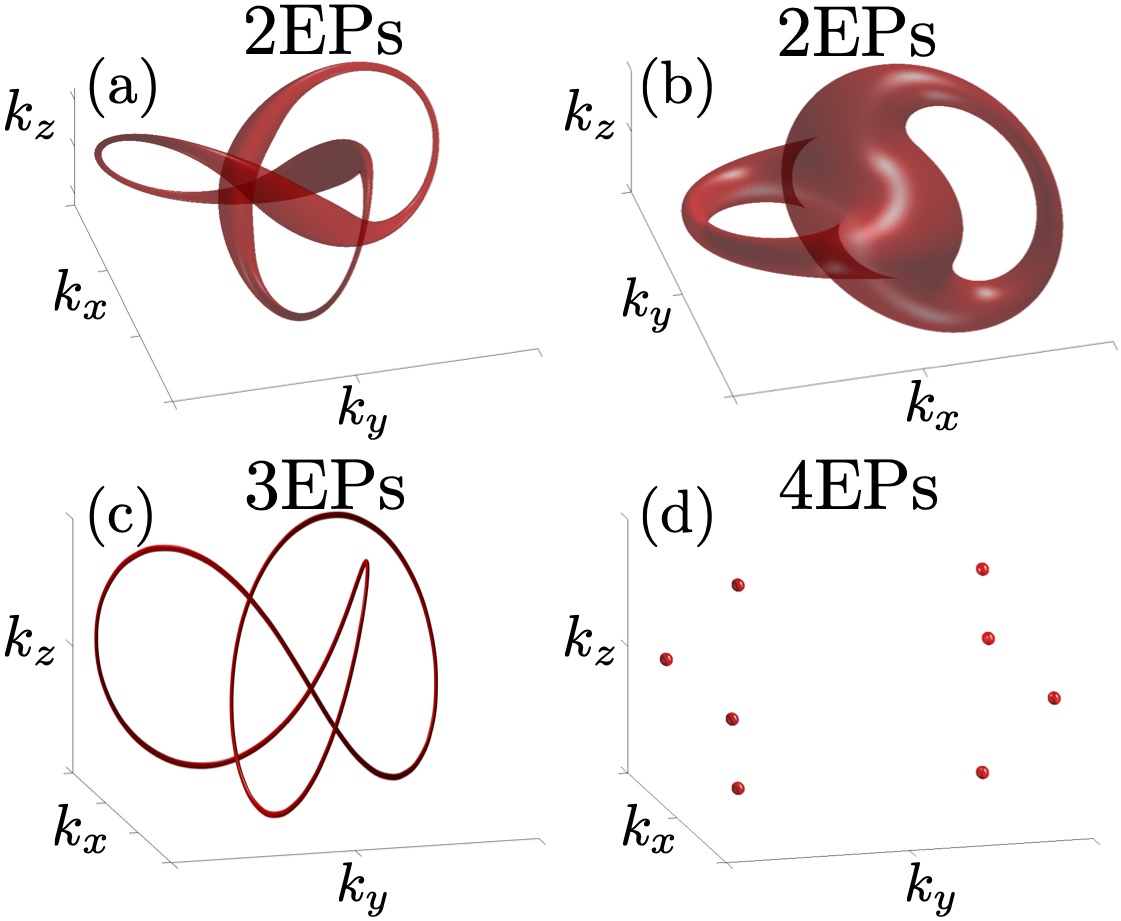}
\caption{$\PT$ symmetry protected exceptional nodal degeneracies in three-dimensional (3D) systems. In (a) and (b), second-order exceptional surfaces of the Hamiltonian in Eq.~\eqref{eq:2bandham} are displayed. (a) An inflated exceptional trefoil knot. (b) An inflated Hopf-link. The surfaces are obtained by adding a symmetry-preserving perturbation to systems hosting ordinary nodal knot degeneracies. The surfaces (a) and (b) are of genus 1 and 2, respectively, and higher genus surfaces can be obtained in a systematic manner. (c) An exceptional trefoil knot of order three is shown, which generically occurs in $\PT$-symmetric models with three or more bands, e.g., Eq.~\eqref{eq:3bandham}. Lastly, exceptional points of order four are displayed in (d). Together, these exceptional structures provide a complete classification of exceptional nodal topologies protected by $\PT$ symmetry in 3D, cf. Table \ref{tab:classification}.}\label{fig:tease}
\end{figure} 

Here, we show that $\PT$ symmetry protects a remarkably rich family of nodal EP topologies in 3D. By explicit construction, we below provide a classification thereof, including possibilities such as knotted surfaces of arbitrary genus of 2EPs, knotted lines of 3EPs and point-like 4EPs, cf. Figure~\ref{fig:tease} and Table~\ref{tab:classification}. All structures are generic and stable in the sense that they are indeed protected by $\PT$ symmetry; small perturbations can destroy the EP topologies only if they break $\PT$. This points to their experimental relevance mainly in photonic systems where the $\PT$ can be realized with great precision and system parameters may be tuned with excellent control \cite{OzRoNoYa2019}. 

\begin{table}[h]
\begin{tabular}{ |c|c|c| }
 \hline
 \multicolumn{3}{|c|}{\bf{Exceptional Nodal Topologies}} \\
 \hline
 Dimension& Without $\PT$ &With $\PT$
 \\
 \hline
 $D=1$  & ---    &2EPs\\
 \hline
 \multirow{2}{*}{$D=2$}&  \multirow{2}{*}{2EPs}  & Lines of 2EPs\\ & &3EPs  
 \\
 \hline
  \multirow{3}{*}{$D=3$} & \multirow{3}{*}{Knots of 2EPs} & Surfaces of 2EPs \\ & & Knots of 3EPs \\ & & 4EPs\\
 \hline
\end{tabular}
\caption{The impact of $\PT$ symmetry on the exceptional nodal topologies in systems of physically relevant dimensions.} \label{tab:classification}
\end{table}

\section{$\PT$ and codimension of $N$EPs}
The codimension of $n$EPs is related to the properties of the corresponding characteristic polynomial $\Ch(\lambda)$. Imposing $\PT$ symmetry yields the following constraints on $\Ch(\lambda)$ \cite{3Tsuneya}:
\begin{align}
\Ch_{\PT}(\lambda) &= \det\left(H_{\PT}-\lambda\right) = \det \left(H_{\PT}^*-\lambda\right) \label{eq:PTcon},
\end{align}
where $^*$ denotes complex conjugation. Explicitly, Eq.~\eqref{eq:PTcon} reads,
\begin{equation}
\Ch_{\PT}(\lambda) = \sum_{j=0}^n a_j\lambda^j = \sum_{j=0}^n a_j^*\lambda^j.
\end{equation}
Consequently, $a_j$ are real-valued functions, indicating that $\PT$ reduces the number of real constraints for $n$EPs to $n-1$, thus requiring the tuning of at least $n-1$ real parameters. Below, we show by explicit construction that these real constraints indeed have non-trivial real-valued solutions in 3D, going beyond earlier works such as Ref.~\cite{3Tsuneya}. The general argument confirms that a further decrease of the codimension of the EPs are not stabilizedby $\PT$ and therefore relies on fine-tuning. Consequently, our findings constitute a classification of EP topologies protected by $\PT$ symmetry in 3D since the exceptional structures map over to embeddings of the respective dimension, ensuring that all corresponding topologies can be obtained \cite{algknot,knotnodal,Milnor}.

Throughout this Letter, we use a representation such that $\PT$ symmetry amounts to complex conjugation, meaning that matrix representations of $\PT$-symmetric Hamiltonians will consist of purely real entries and $(\PT)^2=1$.

\begin{figure*}[!t]
\centering
\includegraphics[width=\textwidth]{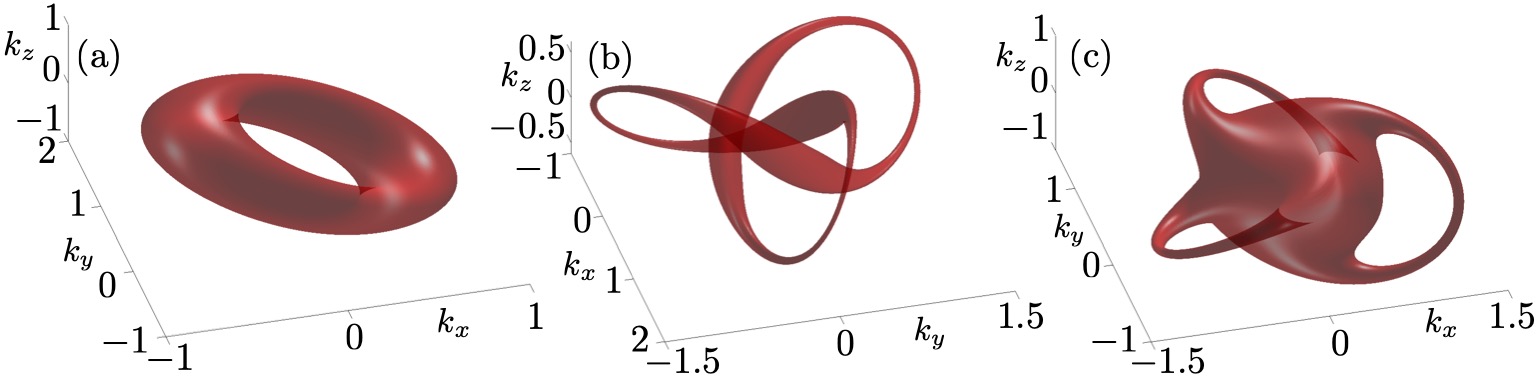}
\caption{Second-order exceptional surfaces of the Hamiltonian in Eq.~\eqref{eq:2bandham}. In all panels, $M=0.5$. In (a), $p=q=1$ and $\delta = 0.4$, resulting in an exceptional torus. In (b), $p=3$, $q=2$ and $\delta=0.15$, resulting in an inflated exceptional trefoil knot. Despite appearing topologically different, both (a) and (b) display genus 1 surfaces, since the inflated trefoil knot is an embedding of the torus. In (c), $p=q=3$ and $\delta=0.4$, resulting in a genus 3 surface. In this controlled manner, exceptional nodal surfaces of any topology can be achieved by choosing $p=q=n$.}\label{fig:surfaces}
\end{figure*}

\section{Exceptional nodal surfaces}
$\PT$ symmetry preserving anti-Hermitian terms split ordinary Hermitian nodal points to exceptional rings in two-band systems \cite{Budich2019}. Generalizing this reasoning to higher dimensions, nodal lines split to exceptional surfaces when exposed to similar perturbations. To demonstrate this, we first show that $\PT$ symmetry-protected nodal torus knots are split into exceptional knotted tori. Consider the following $\PT$-symmetric Hermitian Hamiltonian \cite{hermknot},
\begin{align}
H_{(p,q)} &= \text{Re}\left[\zeta^p(k_x,k_y,k_z)+\xi^q(k_x,k_y,k_z)\right] \sigma^x \nonumber
\\
& + \text{Im}\left[\zeta^p(k_x,k_y,k_z)+\xi^q(k_x,k_y,k_z)\right] \sigma^z,
\end{align}
where a $\sigma^y$-term is absent due to the symmetry. Here, $\zeta(k_x,k_y,k_z)=k_x+i k_y$, $\xi = k_z+i\left[M-(k_x^2+k_y^2+k_z^2)\right]$, $M\in \R$, $\sigma^i$ the Pauli matrices, and $k_i$ the lattice momentum components. Here, $p$ and $q$ are positive integers, and the corresponding nodal points form torus knots when $p$ and $q$ are relatively prime, and torus links otherwise. By adding a constant symmetry preserving anti-Hermitian term, resulting in the following Hamiltonian,
\begin{equation} \label{eq:2bandham}
H = H_{(p,q)}+i\delta \sigma^y,
\end{equation}
for $\delta \in \R$, the nodal knots are inflated to form second-order exceptional surfaces of knotted tori, displayed in Figs.~\ref{fig:tease}(a), \ref{fig:surfaces}(a), and (b). This construction is however not restricted to torus knots, but is also applicable for the hyperbolic Turk's head knots by following the construction in Ref. \cite{ourknots2}. Details can be found in the online Supplemental Material \cite{SuppMat}.

Despite being topologically different, in the sense that the different knotted tori cannot be continuously deformed into each other, they are all different examples of embeddings of the regular torus into $\R^3$. Thus, the genus of a knotted torus is alway $1$---the genus is independent of the embedding. Since embeddings of 2D surfaces in 3D spaces are completely classified by the corresponding genus, we expect higher genus exceptional surfaces to appear as well. To this end, let us again consider the Hamiltonian in Eq.~\eqref{eq:2bandham}, and let $p=q=n$. The exceptional structure for $H$ consist of $n$ individually linked tori. When $\delta$ becomes large enough, the tori are merged together close to the origin of momentum space, and the exceptional structure attain the form of a sphere with $n$ handles, i.e., a genus $n$ surface. This is explicitly shown in Figs.~\ref{fig:tease} (b) and \ref{fig:surfaces} (c) for $n=2$ and $n=3$, respectively, while higher genus examples are provided in the Supplemental Material \cite{SuppMat}. Thus, it is possible to obtain second-order exceptional surfaces with any genus. These surfaces are furthermore generic and stable, in the sense that they are protected by $\PT$ symmetry.

\section{Knotted exceptional lines of order three} 
\begin{figure*}[t]
\centering
\includegraphics[width=\textwidth]{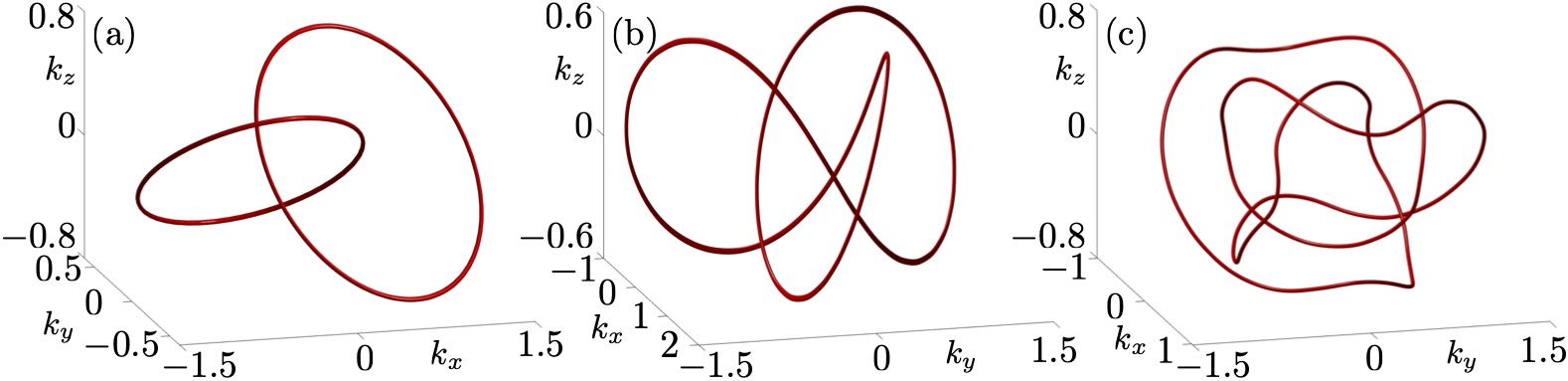}
\caption{Knotted and linked exceptional lines of order three appearing as triple eigenvalue degeneracies in the model described by Eq.~\eqref{eq:3bandham}. For all panels, $\alpha=1$, $\beta=-1$ and $\Lambda=100$. (a) An exceptional Hopf-link, which corresponds to $p=q=2$ in Eqs.~\eqref{eq:f1torus} and \eqref{eq:f2torus}. (b) An exceptional trefoil knot, for which $p=3$ and $q=2$. In both cases, $M=0.5$. The exceptional knot in (c) is a figure-eight knot, appearing when choosing $f_1$ and $f_2$ as in Eqs.~\eqref{eq:f1TH} and \eqref{eq:f2TH}, with $\epsilon=1$. }\label{fig:knots}
\end{figure*}

The plethora of exceptional nodal topologies protected by $\PT$ symmetry does not only contain second-order surfaces, but also higher-order degeneracies. Below, building on the recent examples on 3EPs (points) in two dimensions  \cite{ips3,3Tsuneya}, we exemplify this by explicit construction of a three-band model hosting knotted exceptional lines of order three. Consider the Hamiltonian given by
\begin{equation} \label{eq:3bandham}
H_{3} = \begin{pmatrix} f_2&0&\Lambda\\ \alpha&0&\beta\\f_2&f_1&-f_2 \end{pmatrix},
\end{equation}
where $f_1$ and $f_2$ are continuously differentiable functions of the lattice momentum components, and $\alpha,\beta,\Lambda\in \R$. The corresponding characteristic equation determining the eigenvalues reads,
\begin{align} \label{eq:cp3}
&\quad f_1\left(\alpha\, \Lambda -\beta \,f_2 \right)+ \nonumber
\\
&+\lambda\left(f_2^2+\Lambda \,f_2+\beta\, f_1\right) - \lambda^3=0.
\end{align}
To have third-order exceptional structures, all eigenvalues have to coalesce. This happens exactly when
\begin{align}
 f_1\left(\alpha\, \Lambda -\beta \,f_2 \right)=0, 
 \\
 \left(f_2^2+\Lambda \,f_2+\beta\,f_1\right)=0.
\end{align}
When $|\Lambda|$ is sufficiently large, the only solution becomes $f_1=f_2=0$. Recalling the interpretation of a knot as the intersection of two implicitly defined surfaces, $f_1$ and $f_2$ can be defined such that their common zeros resemble any knot \cite{algknot,knotnodal}. By choosing
\begin{align}
f_1 &= \text{Re}\left[\zeta^p(k_x,k_y,k_z)+\xi^q(k_x,k_y,k_z)\right], \label{eq:f1torus}
\\
f_2 &=\text{Im}\left[\zeta^p(k_x,k_y,k_z)+\xi^q(k_x,k_y,k_z)\right], \label{eq:f2torus}
\end{align}
with $\zeta$ and $\xi$ as above, the exceptional lines are torus knots \cite{Milnor}, while an exceptional hyperbolic figure-eight knot is obtained when
\begin{align}
f_1 &=(k_y^2-k_z^2)\epsilon^2+R\left(8R^2-2\epsilon^2\right), \label{eq:f1TH}
\\
f_2 &=2\sqrt{2}R\,k_y\,k_z + k_x\left(8R^2-2\epsilon^2\right), \label{eq:f2TH}
\end{align}
where $R^2=\epsilon^2-\left(k_x^2+k_y^2+k_z^2\right)$ \cite{frenchknot}. In Fig.~\ref{fig:knots}, a sample of illustrations is provided, including examples from both of the cases above. The topology of the knots can be determined by computing knot invariant polynomials, such as the Alexander polynomial \cite{ourknots2,CH4D}. In the Supplemental Material \cite{SuppMat}, additional examples, including hyperbolic Turk's head knots, are provided.

To emphasize that the knotted exceptional structures are indeed of order three, let us consider the eigenvectors, which are given by
\begin{equation}
V_i = \left(-\frac{\Lambda}{f_2-E_i},-\frac{-\beta\, f_2+\alpha \, \Lambda + \beta\, E_i}{\left(f_2-E_i\right)},1\right)^{\text{T}},
\end{equation}
where $E_i$ denotes the corresponding eigenvalue, and $i=1,2,3$. Taking the difference $V_i-V_j$ for $i\neq j$, and letting $E_i\to E_j$, the eigenvectors indeed coalesce when the eigenvalues do, and hence, the degeneracies at $E_i=0$ are exceptional of order three. They are furthermore stable towards symmetry-preserving perturbations, which follows by the stability of the models in Refs.~\cite{ourknots,ourknots2}.

\section{4EPs in perturbed Dirac semimetals} 
\begin{figure}[!b]
\centering
\includegraphics[width=\columnwidth]{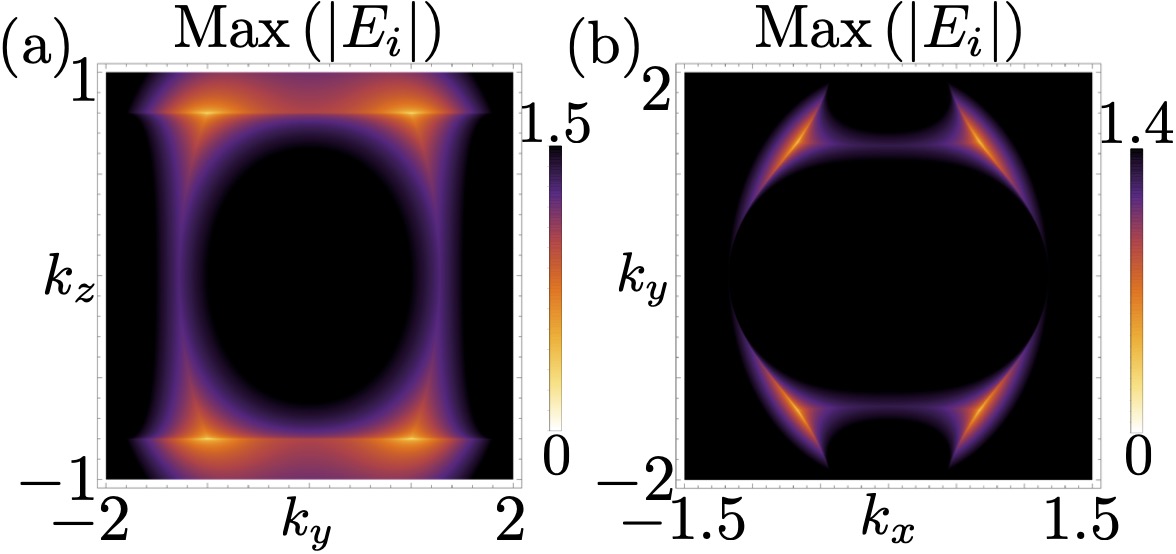}
\caption{4EPs appearing in the system described by the Hamiltonian in Eq.~\eqref{eq:4bandham}. In both panels, $\delta_4=0.8$‚ $\delta_6=1$ and $\delta_{12}=1.1$. In (a), $k_x=0$, and the 4EPs occur at $(k_y,k_z)=(\pm_1\delta_6,\pm_2 \delta_4)$, where $\pm_1$ and $\pm_2$ are independent of each other. In (b), $k_z=0$ and the 4EPs occur at $(k_x,k_y)=(\pm_1 0.669048,\pm_2 1.31681)$. Since 4EPs necessarily occur at $E=0$ they correspond exactly to when the maximum of the absolute value of all eigenvalues vanish simultaneously. These points are furthermore stable towards any small, but finite, perturbation, and are thus generic in systems protected by $\PT$ symmetry.}\label{fig:4EP}
\end{figure}
Finally, we demonstrate the existence of 4EPs in $\mathcal{PT}$-symmetric four-band systems. Let us start with the Hamiltonian of a real Dirac semimetal,
\begin{equation}
H_{\D} = k_x\Gamma^1+k_y\Gamma^2+k_z\Gamma^3,
\end{equation}
where $\Gamma^1=\sigma_1\otimes \tau_0$, $ \Gamma^2 =\sigma_2 \otimes \tau_2$ and $\Gamma^3 = \sigma_3 \otimes \tau_0$, together with $\Gamma^4= \sigma_2\otimes \tau_1$ and $\Gamma^5 = \sigma_2\otimes \tau_3$ constitute a representation of the Clifford algebra. Here, $\sigma_i$ and $\tau_i$ denote the Pauli matrices. Anti-commuting $\Gamma^1,...,\Gamma^4$ provides a complete basis of $4\times 4$-matrices. Adding an anti-Hermitian term on the form $H_{\text{AH}}=i\delta_4 \Gamma^4+\delta_6\Gamma^6 + k_x\delta_{12}\Gamma^{12}$, to obtain
\begin{equation} \label{eq:4bandham}
H= H_{\D}+H_{\text{AH}},
\end{equation}
with $\Gamma^6 =\Gamma^{[1}\Gamma^{2]} $, $\Gamma^{12} =\Gamma^{[1}\Gamma^2\Gamma^{3]} $ and $\delta_4 , \delta_6, \delta_{12} \in \R$, splits the Dirac node at energy $E=0$ into eight 4EPs, all located at $E=0$. Here, $A^{[\mu_1}\cdot...\cdot A^{\mu_n]}=\frac{1}{n!}\epsilon_{\mu_1...\mu_n}A^{\mu_1}\cdot ...\cdot A^{\mu_n}$ with $\epsilon_{\mu_1...\mu_n}$ the Levi-Civita symbol. This can be seen by studying the characteristic polynomial, which attains the form of a depressed quartic. If such polynomials have four-fold degenerate solutions, they are necessarily located at $E=0$. In Fig.~\ref{fig:4EP}, the maximum of the absolute values of all eigenvalues $E_1,...,E_4$ are displayed as a density plot in the appropriate planes in momentum space. The 4EPs occur exactly when the maximum value is zero since then all eigenvalues necessarily coalesce at $E=0$. The points are indeed exceptional since the dimensions of the eigenspaces at those points are one, indicating that the eigenvalues $E_1=E_2=E_3=E_4=0$ share one eigenvector. 

\section{Discussion} 
In this Letter, we provide a classification of exceptional nodal topologies protected by $\PT$ symmetry in 3D. Through explicit construction, we demonstrate the existence of several hitherto overlooked types of exceptional nodal degeneracies: knotted surfaces of arbitrary genus of 2EPs, knotted lines of 3EPs, and 4EPs, thus extending the already rich plethora of NH topological phases \cite{review,Budich2019,Yoshida2019,Okugawa,Zhou,gong,lieu2,Kawabata2019,ZhLe2019,KawabataExceptional,esakisatohasebekohmoto,lee,BBC,yaowang}. We note that higher-order EPs require tuning of more parameters than allowed by dimensional constraints, making the list of exceptional nodal topologies complete for physically relevant systems  if additional tuning variables are not considered.

We emphasize again that our findings rely only on the presence of $\PT$ symmetry. Being arguably the most widely implemented symmetry in photonic experimental setups, make them highly relevant in optics \cite{OzRoNoYa2019}. It should be possible to experimentally simulate the models we present using single-photon interferometry as in Ref.~\cite{knotexp}. We furthermore note that 2EPs, together with their bulk Fermi arcs, as well as the more exotic knotted lines of 2EPs and their concomitant Fermi-Seifert surfaces were both experimentally realized and observed shortly after being theoretically predicted \cite{NHarc,knotexp}. This suggests that topological phases of higher-order EPs are within experimental reach in several platforms and encourages the search for tailor-made setups where they can be realized.

\emph{Acknowledgments.---} We are grateful to Lukas R\o dland and Gregory Arone for fruitful discussions. We also thank Jan Budich, Ipsita Mandal and Johan Carlstr\"om for related collaborations. The authors are supported by the Swedish Research Council (VR) and the Wallenberg Academy Fellows program as well as the project Dynamic Quantum Matter of the Knut and Alice Wallenberg Foundation.

\emph{Note added:} An independent work shows how symmetry-protected 4EPs may naturally be realized in three-dimensional correlated many-body systems \cite{note}.

 \newpage
\begingroup
\onecolumngrid
\appendix
\section{Supplementary Material for ``Classification of Exceptional Nodal Topologies Protected by $\mathcal{PT}$ Symmetry''}

In this supplementary material, we provide details on our derivations and arguments presented in the main text, along with complementary figures. 

\subsection{Hyperbolic nodal knots}
Here, we provide a short summary on how to construct functions whose common zeros represent hyperbolic Turk's head knots. For further details, see Ref. \onlinecite{ourknots2}.

In contrast to torus knots, which are described as zeros of a complex polynomial \cite{Milnor}, the Turk's head knots are illustrated as the zeros of a real polynomial, constrained on $S^3_{\epsilon}$ (the three-sphere of radius $\epsilon$) \cite{frenchknot}. Explicitly, the polynomial is,
\begin{equation} \label{eq:hyp1}
F\left\{x,y,\text{Re}\left[(z+it)^N\right],\text{Im}\left[(z+it)^N\right]\right\},
\end{equation}
with
\begin{align} \label{eq:hyp2}
F(x,y,z,t)&=\left\{z\left[x^2+y^2+z^2+t^2\right]+x\left[8x^2-2\left(x^2+y^2+z^2+t^2\right)\right], \right. \nonumber
\\
&\quad \left. \sqrt{2}t\,x+y\left[8x^2-\left(x^2+y^2+z^2+t^2\right)\right]\right\},
\end{align}
where $x,y,z,t$ denote coordinates on $\R^4$ and $N\in \mathbb{N}$. Note that $F:\R^4\to \R^2$. By finding the zeros, restricting them to $S^3_{\epsilon}$, and introducing the lattice momentum appropriately, these knots can be realized as third-order exceptional degeneracies. In Figure~\ref{fig:hypknots}, this is illustrated for different values of $N$, with the lattice momentum introduced as,
\begin{equation}
x=\epsilon^2-\left(k_x^2+k_y^2+k_z^2\right), \quad y=k_x, \quad z=k_y, \quad t=k_z.
\end{equation}
Additionally, these hyperbolic knots can be inflated to form knotted surfaces, just as displayed for the torus knots in the main text. This is illustrated in Figure~\ref{fig:compknots}, along with a more complicated inflated torus knot.

\subsection{Higher genus surfaces}
\begin{figure*}[t]
\centering
\includegraphics[width=\textwidth]{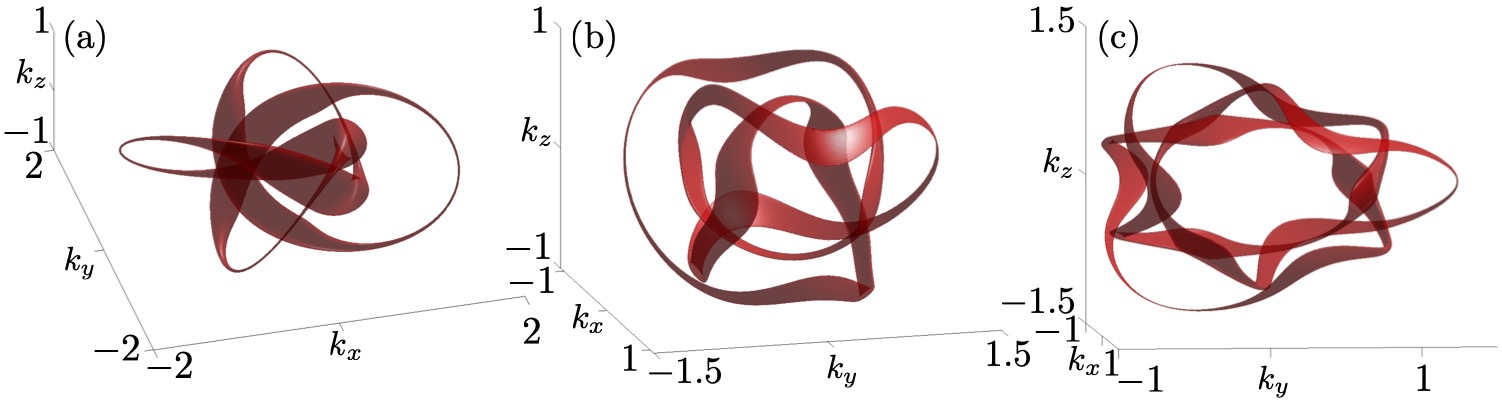}
\caption{Exceptional knotted surfaces of order 2. (a) displays an inflated $(4,3)$ torus knot, obtained from the Hamiltonian in Eq.~\eqref{eq:2bandham}, with $p=4$, $q=3$, $M=1$ and $\delta=0.35$. In (b) and (c), an inflated figure-eight knot and inflated Borromean rings are shown, obtained by using Eqs.~\eqref{eq:hyp1} and \eqref{eq:hyp2} in the Hamiltonian Eq.~\eqref{eq:2bandham}, with $\delta=0.15$, $\epsilon=1$ and $N=2,3$ respectively.}\label{fig:compknots}
\end{figure*} 
\begin{figure*}[t]
\centering
\includegraphics[width=\textwidth]{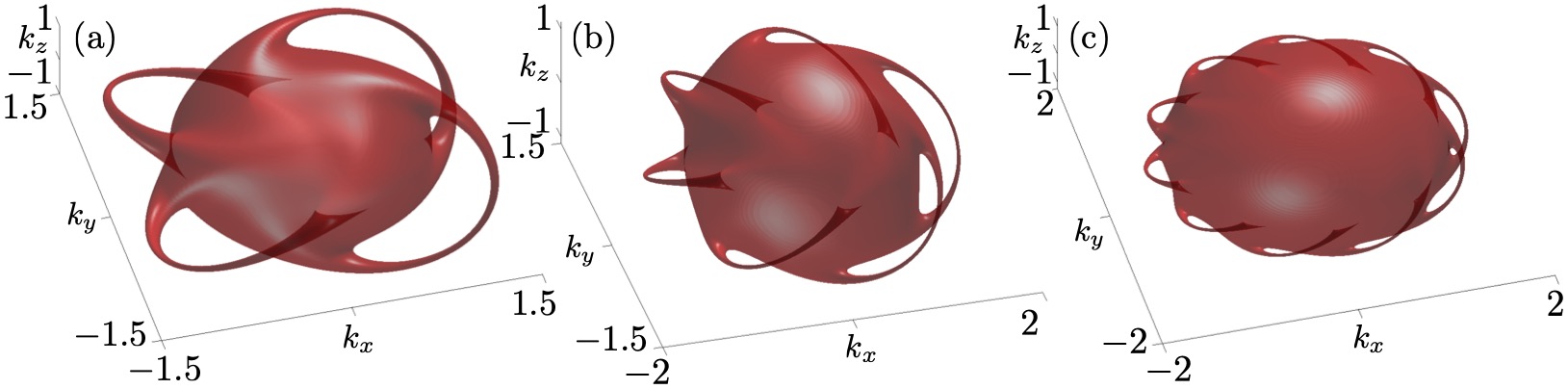}
\caption{higher genus exceptional surfaces of order 2 from the Hamiltonian Eq.~\eqref{eq:2bandham}. In all panels, $M=1$. In (a), $p=q=4$, and $\delta=1$, giving a genus 4 surface. In (b), $p=q=6$, and $\delta=2.5$ resulting in a genus 6 surface. Lastly, (c) displays a genus 10 surface, obtained by choosing $p=q=10$ and $\delta=30$.}\label{fig:highergenus}
\end{figure*} 
As mentioned in the main text, the presented techniques can in principle be used to achieve second-order exceptional surfaces of arbitrary genus. Here, we present a few additional figures to strengthen this claim. This is done by using the Hamiltonian on the form,
\begin{equation}
H_2 =  \text{Re}\left[\zeta^p(k_x,k_y,k_z)+\xi^q(k_x,k_y,k_z)\right] \sigma^x + \text{Im}\left[\zeta^p(k_x,k_y,k_z)+\xi^q(k_x,k_y,k_z)\right] \sigma^z + i\delta \sigma^y,
\end{equation}
taking $p=q=n$, and $\delta\in \R$. In Figure~\ref{fig:highergenus}, we illustrate this for $n=4,6,10$, giving exceptional surfaces of genus 4, 6 and 10 respectively.

\subsection{Controlling topology of third-order exceptional lines}
\begin{figure*}[t]
\centering
\includegraphics[width=\textwidth]{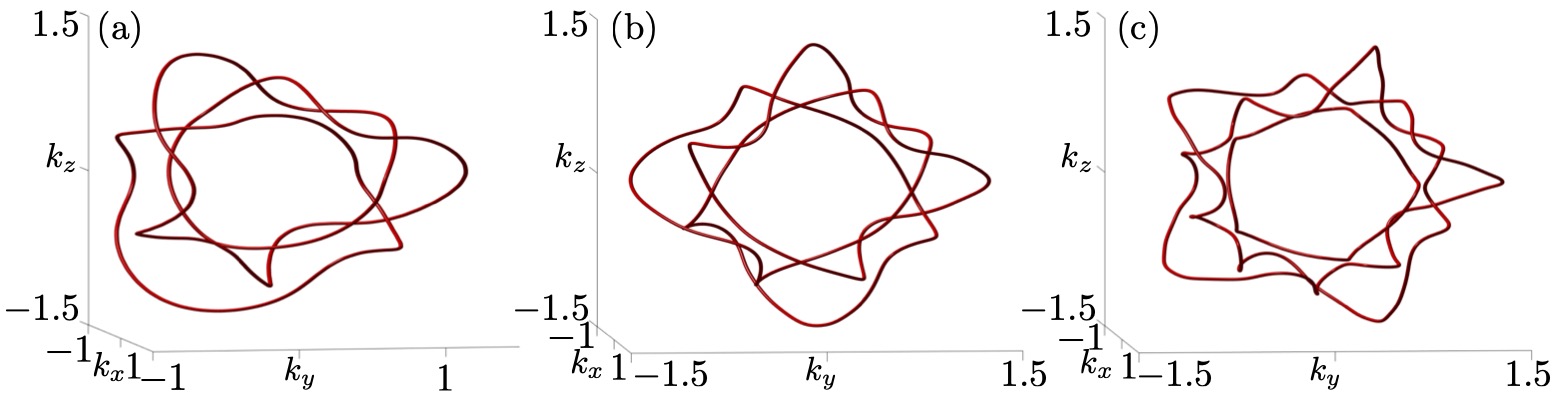}
\caption{Hyperbolic exceptional knots of order 3 of the Hamiltonian Eq.~\eqref{eq:3bandham}, when defining $f_1$ and $f_2$ in terms of the components of the polynomial $F$, defined in Eqs.~\eqref{eq:hyp1} and \eqref{eq:hyp2}. For all panels, $\alpha=1$, $\beta=-1$ and $\Lambda=100$, while $N=3,4,5$ in (a), (b) and (c) respectively. (a) displays Borromean rings, while (b) and (c) are mote complicated hyperbolic knots in the family of Turk's head knots.}\label{fig:hypknots}
\end{figure*} 

Consider an arbitrary real $3\times 3$ Hamiltonian,
\begin{equation}
H_3 = \begin{pmatrix}a_1&a_2&a_3\\a_4&a_5&a_6\\a_7&a_8&a_9 \end{pmatrix}
\end{equation}
with all $a_j$ being real-valued, continuously differentiable function of the lattice momentum components, $j=1,...,9$. Since we are interested in when all the eigenvalues of $H_3$ coalesce, we can assume that $H_3$ is traceless and take $a_5=0$, $a_1=-a_9$. The corresponding characteristic polynomial reads,
\begin{equation}
\Ch_{H_3}(\lambda) = a_2a_6a_7+a_3a_4a_8-a_1a_6a_8+a_1a_2a_4+\lambda\left(a_2a_4+a_3a_7+a_6a_8-a_1^2\right)-\lambda^3
\end{equation}
Triple eigenvalue degeneracies occur when,
\begin{align}
a_2a_6a_7+a_3a_4a_8-a_1a_6a_8+a_1a_2a_4&=0,
\\
a_2a_4+a_3a_7+a_6a_8-a_1^2&=0.
\end{align}
Taking $a_2=0$ gives us
\begin{align}
a_8\left(a_3a_4-a_1a_6\right)&=0, \label{eq:supptrip1}
\\
a_3a_7+a_6a_8-a_1^2 &=0. \label{eq:supptrip2}
\end{align}
We note that Eq.~\eqref{eq:supptrip1} is satisfied if either $a_8=0$ or $a_3a_4-a_1a_6=0$. Inspired by Ref. \onlinecite{ourknots}, we let $|a_3|\gg0$, and thus forcing $a_8=0$ in order to Eq.~\eqref{eq:supptrip1} to hold. By furthermore letting $a_1=a_7$, Eqs.~\eqref{eq:supptrip1} and \eqref{eq:supptrip2} hold when $a_1=a_8=0$. Recalling the constructions in Refs. \onlinecite{ourknots,ourknots2}, we note that by letting the common zeros of $a_1$ and $a_8$ represent knots, the resulting exceptional line will indeed be knotted. Let $f_1$ and $f_2$ denote such functions, define $a_1:=f_2$, $a_8=f_1$, and let $a_3=\Lambda$, $a_4=\alpha$ and $a_6=\beta$, with $\alpha,\beta,\Lambda\in \R$, and $|\Lambda|\gg0$. Then, the resulting Hamiltonian,
\begin{equation}
H_3 = \begin{pmatrix} f_2&0&\Lambda\\ \alpha&0&\beta\\ f_2&f_1&-f_2\end{pmatrix}
\end{equation}
host third-order exceptional knots. Torus knots are obtained when, e.g., $f_1=\text{Re}\left[\zeta^p(k_x,k_y,k_z)+\xi^q(k_x,k_y,k_z)\right]$ and $f_2=\text{Im}\left[\zeta^p(k_x,k_y,k_z)+\xi^q(k_x,k_y,k_z)\right]$, as discussed in the main text, while hyperbolic Turk's head knots are obtained by choosing $f_1$ and $f_2$ as the first and second components of the function $F$ defined in Eqs.~\eqref{eq:hyp1} and \eqref{eq:hyp2}.


\begin{thebibliography}{10}

\bibitem{benderboettcher}
C.M. Bender and S. Boettcher, {\em Real Spectra in Non-Hermitian Hamiltonians Having $\mathcal{P} \mathcal{T}$ Symmetry}, \href{https://journals.aps.org/prl/abstract/10.1103/PhysRevLett.80.5243}{Phys. Rev. Lett. {\bf 80}, 5243 (1998)}.

\bibitem{bender}
C.M. Bender, {\em Making sense of non-Hermitian Hamiltonians}, \href{http://iopscience.iop.org/article/10.1088/0034-4885/70/6/R03/meta}{Rep. Prog. Phys. {\bf 70}, 947 (2007)}.

\bibitem{NHbook}
Yuto Ashida, Zongping Gong, Masahito Ueda, {\em Non-Hermitian Physics}, \href{https://dx.doi.org/10.1080/00018732.2021.1876991}{Advances in Physics 69, 3 (2020)}.

\bibitem{OzRoNoYa2019} S. K. \"{O}zdemir, S. Rotter, F. Nori, and L. Yang, \textit{Parity-time symmetry and exceptional points in photonics}, \href{https://www.nature.com/articles/s41563-019-0304-9}{Nat. Mater. \textbf{18}, 783-798 (2019)}.
  
\bibitem{EPreview}
M.-A. Miri and A. Alu,  {\em Exceptional points in optics and photonics}, \href{https://science.sciencemag.org/content/363/6422/eaar7709}{Science {\bf 363}, eaar7709 (2019)}.
 
\bibitem{hasankane}
M.Z. Hasan and C.L. Kane, {\em Colloquium: Topological insulators}, \href{https://journals.aps.org/rmp/abstract/10.1103/RevModPhys.82.3045}{Rev. Mod. Phys. {\bf 82}, 3045 (2010)}.

\bibitem{qizhang}
X.-L. Qi and S.-C. Zhang, {\em Topological insulators and superconductors}, \href{https://journals.aps.org/rmp/abstract/10.1103/RevModPhys.83.1057}{Rev. Mod. Phys. {\bf 83}, 1057 (2011)}.

\bibitem{review}
 E.J. Bergholtz, J.C. Budich, and F.K. Kunst, {\em Exceptional topology of non-Hermitian systems}, \href{https://journals.aps.org/rmp/abstract/10.1103/RevModPhys.93.015005}{Rev. Mod. Phys. {\bf 93}, 015005 (2021)}.
 
 \bibitem{NHarc}
L. Lu, Z. Wang, D. Ye, L. Ran, L. Fu, J. D. Joannopoulos, and M. Solja\v{c}i\'{c}, {\em Observation of bulk Fermi arc and polarization half charge from paired exceptional points}, \href{https://science.sciencemag.org/content/359/6379/1009.abstract}{Science {\bf 359}, 1009 (2018)}.
 
 \bibitem{BerryDeg}
M. Berry,  {\em Physics of Nonhermitian Degeneracies}, \href{https://link.springer.com/article/10.1023\%2FB\%3ACJOP.0000044002.05657.04}{Czechoslovak Journal of Physics {\bf 54}, 1039 (2004)}.

\bibitem{Heiss}
W. D. Heiss, {\em The physics of exceptional points}, \href{http://adsabs.harvard.edu/abs/2012JPhA...45R4016H}{Journal of Physics A. {\bf 45}, 444016 (2012)}.

\bibitem{koziifu}
V. Kozii and L. Fu, {\em Non-Hermitian Topological Theory of Finite-Lifetime Quasiparticles: Prediction of Bulk Fermi Arc Due to Exceptional Point}, \href{https://arxiv.org/abs/1708.05841}{arXiv:1708.05841 (2017)}.

\bibitem{EPrings}
Y. Xu, S.-T. Wang, and L.-M. Duan, {\em Weyl Exceptional Rings in a Three-Dimensional Dissipative Cold Atomic Gas}, \href{https://doi.org/10.1103/PhysRevLett.118.045701}{Phys. Rev. Lett. {\bf 118}, 045701 (2017)}.

\bibitem{CaBe2018} J. Carlstr\"{o}m and E. J. Bergholtz, \textit{Exceptional links and twisted Fermi ribbons in non-Hermitian systems}, \href{https://journals.aps.org/pra/abstract/10.1103/PhysRevA.98.042114}{Phys. Rev. A \textbf{98}, 042114 (2018)}.

\bibitem{EPringExp}
A. Cerjan, S. Huang, K. P. Chen, Y. Chong, and M. C. Rechtsman, {\em Experimental realization of a Weyl exceptional ring}, \href{https://www.nature.com/articles/s41566-019-0453-z}{Nat. Photon. {\bf 13}, 623 (2019)}.

\bibitem{ourknots}
J. Carlstr{\"o}m, M. St\aa lhammar, J.C. Budich, and E.J. Bergholtz, {\em Knotted Non-Hermitian Metals}, \href{https://journals.aps.org/prb/abstract/10.1103/PhysRevB.99.161115}{Phys. Rev. B {\bf 99}, 161115(R) (2019)}.

\bibitem{ourknots2}
M. St\aa lhammar, L. R\o dland, G. Arone, J.C. Budich and E.J. Bergholtz, {\em Hyperbolic nodal band structures and knot invariants}, \href{https://www.nature.com/articles/s42005-021-00535-1}{SciPost Physics {\bf 7}, 019 (2019)}.

\bibitem{chinghuaknot} X. Zhang, G. Li, Y. Liu, T. Tai, R. Thomale, and C.H. Lee, \textit{Tidal surface states as fingerprints of non-Hermitian nodal knot metals}, \href{https://www.nature.com/articles/s42005-021-00535-1}{Communications Physics {\bf 4}, 47 (2021)}.

\bibitem{yangknot} Z. Yang, C.-K. Chiu, C. Fang, and J. Hu, \textit{Jones Polynomial and Knot Transitions in Hermitian and non-Hermitian Topological Semimetals}, \href{https://journals.aps.org/prl/abstract/10.1103/PhysRevLett.124.186402}{Phys. Rev. Lett. 124, 186402 (2020)}.

\bibitem{Budich2019}
J. C. Budich, J. Carlstr\"om, F. K. Kunst, and E. J. Bergholtz, {\em Symmetry-protected nodal phases in non-Hermitian systems}, \href{https://journals.aps.org/prb/abstract/10.1103/PhysRevB.99.041406}{Phys. Rev. B {\bf{99}}, 041406(R) (2019)}.

\bibitem{Yoshida2019}
T. Yoshida, R. Peters, N. Kawakami, and Y. Hatsugai, {\em Symmetry-protected exceptional rings in two-dimensional correlated systems with chiral symmetry}, \href{https://journals.aps.org/prb/abstract/10.1103/PhysRevB.99.121101}{Phys. Rev. B {\bf{99}}, 121101(R)  (2019)}.

\bibitem{Okugawa}
R. Okugawa and T. Yokoyama, {\em Topological exceptional surfaces in non-Hermitian systems with parity-time and parity-particle-hole symmetries}, \href{https://journals.aps.org/prb/abstract/10.1103/PhysRevB.99.041202}{Phys. Rev. B {\bf  99}, 041202(R) (2019)}.

\bibitem{Zhou}
H. Zhou, J. Y. Lee, S. Liu, and B. Zhen, {\em Exceptional Surfaces in PT-Symmetric Photonic Systems}, \href{https://arxiv.org/abs/1810.06549}{arXiv:1810.06549 (2018)}.

\bibitem{HCphot}
Alexander Szameit, Mikael C. Rechtsman, Omri Bahat-Treidel, and Mordechai Segev, {\em $\PT$-symmetry in honeycomb photonic lattices}, \href{https://journals.aps.org/pra/abstract/10.1103/PhysRevA.84.021806}{Phys. Rev. A {\bf 84}, 021806(R) (2011)}.

\bibitem{Etorus}K. Kimura, T. Yoshida, and N. Kawakami, {\em Chiral-symmetry protected exceptional torus in correlated nodal-line semimetals},
\href{https://journals.aps.org/prb/abstract/10.1103/PhysRevB.100.115124}{Phys. Rev. B {\bf 100}, 115124 (2019)}.

\bibitem{3Tsuneya}P. Delplace, T. Yoshida, and Y. Hatsugai, {\em Symmetry-protected higher-order exceptional points and their topological characterization},
\href{https://arxiv.org/abs/2103.08232}{arXiv:2103.08232}.

\bibitem{ips3} I. Mandal, and E.J. Bergholtz, {\em Symmetry and Higher-Order Exceptional Points}, \href{https://arxiv.org/abs/2103.15729}{arXiv:2103.15729}.

\bibitem{Milnor}
J. Milnor, {\em Singular Points of Complex Hypersurfaces}, \href{https://www.cambridge.org/core/journals/proceedings-of-the-edinburgh-mathematical-society/article/john-milnor-singular-points-of-complex-hypersurfaces-annals-of-mathematics-studies-no-61-princeton-university-press-london-oxford-university-press-1969-30s/E3F3A9C0FF2BCF2D7556AFA185C77BFA}{Princeton University Press (1968)}.

\bibitem{algknot}
S. Akbulut, and H. King, {\em All knots are algebraic}, \href{https://doi.org/10.1007/BF02566217}{Comment. Math. Helvetici {\bf 56} (1981), 339-351}.

\bibitem{knotnodal}
B. Bode, and M. R. Dennis, {\em Constructing a polynomial whose nodal set is any prescribed knot or link}, \href{https://arxiv.org/abs/1612.06328}{arXiv:1612.06328}.

\bibitem{hermknot} R. Bi, Z. Yan, L. Lu, and Z. Wang, {\em Nodal-knot semimetals}, \href{https://journals.aps.org/prb/abstract/10.1103/PhysRevB.96.201305}{Phys. Rev. B {\bf 96}, 201305(R) (2017)}.

\bibitem{SuppMat} See online Supplementary Material for further mathematical details and additional figures.

\bibitem{frenchknot}
B. Perron, {\em Le Noeud Huit est Algebrique R\'eel}, \href{https://link.springer.com/article/10.1007/BF01396628}{Invent. Math. {\bf 65}, p. 441-451 (1982)}.

\bibitem{CH4D}
L. Li, C.H. Lee, and J. Gong, {\em Emergence and full 3D-imaging of nodal boundary Seifert surfaces in 4D topological matter}, \href{https://www.nature.com/articles/s42005-019-0235-4}{Communications Physics {\bf 2} 136 (2019)}.

\bibitem{gong}
Z. Gong, Y. Ashida, K. Kawabata, K. Takasan, S. Higashikawa, and M. Ueda, {\em Topological Phases of Non-Hermitian Systems}, \href{https://journals.aps.org/prx/abstract/10.1103/PhysRevX.8.031079}{Phys. Rev. X {\bf 8}, 031079 (2018)}.


\bibitem{lieu2}
S. Lieu, {\em Topological symmetry classes for non-Hermitian models and connections to the bosonic Bogoliubov--de Gennes equation}, \href{https://link.aps.org/doi/10.1103/PhysRevB.98.115135}{Phys. Rev. B {\bf 98}, 115135 (2018)}.

\bibitem{Kawabata2019}
K. Kawabata, K. Shiozaki, M. Ueda, and M. Sato, {\em Symmetry and Topology in Non-Hermitian Physics},
\href{https://journals.aps.org/prx/abstract/10.1103/PhysRevX.9.041015}{Phys. Rev. X {\bf{9}}, 041015 (2019)}.

\bibitem{ZhLe2019} H. Zhou and J. Y. Lee, \textit{Periodic table for topological bands with non-Hermitian symmetries}, \href{https://journals.aps.org/prb/abstract/10.1103/PhysRevB.99.235112}{Phys. Rev. B \textbf{99}, 235112 (2019)}.

\bibitem{KawabataExceptional} K. Kawabata, T. Bessho, and M. Sato, \textit{Classification of Exceptional Points and Non-Hermitian Topological Semimetals}, \href{https://journals.aps.org/prl/abstract/10.1103/PhysRevLett.123.066405}{Phys. Rev. Lett. {\bf 123}, 066405 (2019)}.

\bibitem{esakisatohasebekohmoto}
K. Esaki, M. Sato, K. Hasebe, and M. Kohmoto, {\em Edge states and topological phases in non-Hermitian systems}, \href{https://journals.aps.org/prb/abstract/10.1103/PhysRevB.84.205128}{Phys. Rev. B {\bf 84}, 205128 (2011)}.


\bibitem{lee}
T.E. Lee, {\em Anomalous Edge State in a Non-Hermitian Lattice}, \href{https://journals.aps.org/prl/abstract/10.1103/PhysRevLett.116.133903}{Phys. Rev. Lett. {\bf 116}, 133903 (2016)}.

\bibitem{BBC}
F.K. Kunst, E. Edvardsson, J.C. Budich, and E.J. Bergholtz, {\em Biorthogonal Bulk-Boundary Correspondence in non-Hermitian Systems}, \href{https://journals.aps.org/prl/abstract/10.1103/PhysRevLett.121.026808}{Phys. Rev. Lett. {\bf 121}, 026808 (2018)}.

\bibitem{yaowang}
S. Yao and Z. Wang, {\em Edge states and topological invariants of non-Hermitian systems}, \href{https://doi.org/10.1103/PhysRevLett.121.086803}{Phys. Rev. Lett. {\bf 121}, 086803 (2018)}.

\bibitem{knotexp}
K. Wang, L. Xiao, J.C. Budich, W. Yi, and P. Xue, {\em Simulating Exceptional Non-Hermitian Metals with Single-Photon Interferometry}, \href{https://arxiv.org/abs/2011.01884}{arXiv:2011.01884}.

\bibitem{note}
L. Crippa, J.C. Budich, and G. Sangiovanni, {\em Fourth-Order Exceptional Points in Correlated Quantum Many-Body Systems}, \href{https://arxiv.org/abs/2106.11987}{arXiv:2106.11987}.

\end{thebibliography}
\end{document}